\documentclass[preprint,
amsmath,amssymb,aps,nofootinbib, Letter]{revtex4-1}

\usepackage{graphicx}
\usepackage{dcolumn}
\usepackage{bm}
\usepackage{color}

\usepackage{multirow}
\usepackage{longtable}
\usepackage{xcolor}
\usepackage{colortbl,booktabs}

\newcommand{\la}{\langle}
\newcommand{\ra}{\rangle}
\newcommand{\qbar}{\bar{q}}
\newcommand{\ubar}{\bar{u}}

\newcommand{\sbar}{\overline{s}}
\newcommand{\cbar}{\overline{c}}

\def\B{{\cal B}}
\newcommand{\SU}{{\text{SU}}}

\def\be{\begin{eqnarray}}
\def\en{\end{eqnarray}}

\begin{document}

\title{Hadronic weak decays of the charmed baryon $\Omega_c$}

\author{ Shiyong Hu, Guanbao Meng, Fanrong Xu\footnote{fanrongxu@jnu.edu.cn}}
\affiliation{
 Department of Physics, Jinan University,
 Guangzhou 510632, People's Republic of China
}%

\bigskip
\begin{abstract}
\bigskip
Two-body hadronic weak decays of the charmed baryon $\Omega_c$, including Cabibbo-favored (CF), singly Cabibbo-suppressed (SCS) and doubly Cabibbo-suppressed (DCS) modes, are studied systematically  in this work.
To estimate nonfactorizable contributions, we work in the pole model for the $P$-wave amplitudes and current algebra
for the $S$-wave amplitudes.  
Among all the channels decaying into a baryon octet and a pseudoscalar meson,
$\Omega_c\to \Xi^0\overline{K}^0$ is the only allowed  CF mode. 
The predicted branching fraction  of order  $3.8\%$ and large and positive decay asymmetry of order $0.50$ 
indicate that a measurement of this mode in the near future is promising.
Proceeding through purely nonfactorizable contributions, the SCS mode $\Omega_c\to\Lambda^0\overline{K}^0$ and DCS mode $\Omega_c\to\Lambda^0 \eta$  are predicted to have branching fractions as large as $0.8\%$
and $0.4\%$, respectively.
The two DCS modes $\Omega_c\to\Sigma^0\eta$  and $\Omega_c\to\Lambda^0\pi^0$ 
are suggested to  serve as  new physics searching  channels for  their vanishing SM background.
\end{abstract}


\maketitle

\section{Introduction}\label{sec:Intro}

The $\Omega_c$ baryon comprises a combination of a charm quark and two strange quarks. 
Classified by the \SU(3) flavor symmetry, the $\Omega_c$ is one of the sextet charmed baryons. 
It is also the heaviest one with a mass
\cite{Tanabashi:2018oca}
\begin{equation}
m=(2695.2\pm 1.7) \, \text{MeV},
\end{equation}
averaged over Belle\cite{Solovieva:2008fw}, CLEO\cite{CroninHennessy:2000bz} and E687\cite{Frabetti:1994dp} experiments.

Recently there have been some experimental progresses on the study of $\Omega_c$.
In 2017, LHCb observed
five new narrow states of $\Omega_c$ via the decay channel $\Xi_c^+ K^-$, 
while the light charmed baryon
$\Xi_c^+$ was reconstructed in the mode $p K^- \pi^+$ \cite{Aaij:2017nav}.
Later in the same year,  four of the five resonances were confirmed independently
by Belle \cite{Yelton:2017qxg}. 
Then various theoretical models have been proposed to interpret these new resonances, including 
excited  states of $\Omega_c$ baryon \cite{Cheng:2017ove, Agaev:2017jyt}, pentaquark picture \cite{Huang:2017dwn,An:2017lwg} and so on. 
The understanding of these $\Omega_c$ resonances is still an open question. 
Progress has also been made 
in the ground state $\Omega_c$. 
In 2018, LHCb reported a measurement of
$\Omega_c$ lifetime \cite{Aaij:2018dso}, 
\begin{equation}
\tau=(2.68\pm 0.26)\times 10^{-13}\, \text{s},\label{eq:lifetime}
\end{equation}
which is nearly four times larger than previous 
world-average value $(69\pm 12)$ fs \cite{Link:2003nq,Adamovich:1995pf,Frabetti:1995bi},
and is consistent with a recent theoretical prediction\cite{Cheng:2018rkz}.

It is well known that the decay of $\Omega_c$ is through weak interactions,  which is distinct  from other 
sextet charmed
baryons 
 as well as $\Omega_c$ excited states
. However, 
about its weak decay information we know much less than 
antitriplet charmed baryons
$(\Lambda_c^+, \Xi_c^0, \Xi_c^+)$  due to its low production.
Though  no absolute branching fraction of $\Omega_c$ has been measured so far,
a ratio between the modes decaying into
$\Xi^0\overline{K}^0$ and $\Omega^- \pi^+$,
\begin{equation}
\frac{\mathcal{B}(\Omega_c^0\to \Xi^0 \overline{K}^0)}{\mathcal{B}(\Omega_c^0\to\Omega^- \pi^+)}
=1.64\pm 0.26\pm 0.12, \label{eq:ratio}
\end{equation}
is reported by Belle
using $980\, \text{fb}^{-1}$ of $e^+ e^-$ annihilation data \cite{Yelton:2017uzv},
while the latter one is taken as a benchmark channel.
As for the channels containing $\Omega^-$ in final states, the
semileptonic decay $\Omega_c^-\to \Omega^- e^+ \nu_e$ was firstly observed 
by CLEO in 2012 
 \cite{Ammar:2002pf}.
 With data accumulated both in Belle-II and LHCb more interesting experimental results are anticipated.
 Hence a theoretical study on weak decays of $\Omega_c$ is necessary and timely.

The theoretical studies of hadronic weak decays of  the $\Omega_c$ baryon  had a long history for
several decades, and 
a fast development in the 
1990s  \cite{Korner:1992wi, Xu:1992sw, Cheng:1993gf, Ivanov:1997ra}. 
It has been widely accepted that nonfactorizable contributions to decay amplitudes play 
an important role in the hadronic decays. 
Various methodologies were developed  to describe the nonfactorizable contributions in charmed baryon decays, including the covariant confined quark model \cite{Korner:1992wi,Ivanov:1997ra}, the pole model \cite{Cheng:1993gf} and
the pole model associated with current algebra \cite{Cheng:1993gf}. 
Some new efforts have also been made in recent years \cite{Dhir:2015tja, Zhao:2018zcb}.
In this work, we shall focus on $1/2^+\to 1/2^++0^-$ decays of $\Omega_c$.

In the pole model, nonfactorizable $S$- and $P$-wave amplitudes for $1/2^+\to 1/2^++0^-$ decays 
are dominated by $1/2^-$ low-lying baryon resonances and $1/2^+$ ground-state baryon poles, respectively. 
The estimation of pole amplitudes is a challenging and nontrivial
task since  weak baryon matrix elements and strong
coupling constants of ${1\over 2}^+$ and ${1\over 2}^-$ baryon
states are involved.  As a consequence, the nonfactorizable contribution evaluated from pole diagrams,  is far more uncertain than the factorizable terms. 
The difficulty emerges in particular  for $S$-wave terms as
they require the information of the troublesome negative-parity baryon resonances which are not well understood in the quark model. 
Fortunately, the trick of current algebra helps to avoid evaluation involving ${1\over 2}^-$ state in 
the soft-meson limit \cite{Brown:1966zz, Gronau:1972pj, Cheng:1991sn, Cheng:2018hwl, Zou:2019kzq}.
Although the pseudoscalar meson produced in $\B_c\to \B+P$ decays is in general not truly soft, current algebra approach seems to work empirically well for $\Lambda_c^+\to \B+P$ decays \cite{Cheng:2018hwl,Zou:2019kzq}. Moreover, the predicted negative decay asymmetries by current algebra for both $\Lambda_c^+\to \Sigma^+\pi^0$ and $\Sigma^0\pi^+$ agree in sign with the recent BESIII measurements \cite{Ablikim:2019zwe} (again see \cite{Cheng:2018hwl,Zou:2019kzq} for details). In contrast, the pole model or the covariant quark model and its variant always leads to a positive decay asymmetry for aforementioned two modes. Therefore,
in this work we shall follow \cite{Cheng:2018hwl,Zou:2019kzq,Cheng:2020wmk} to work out
the nonfactorizable $S$-wave amplitudes in $\Omega_c$ decays using current algebra and the nonfactorizable contributions, including $W$-exchange  contribution, to $P$-wave ones using the pole model.

This paper is organized as follows.  In Sec. II we set up the framework for the analysis of
hadronic weak decays of the singly charmed baryon $\Omega_c$, including the topological diagrams and
the formalism for describing factorizable and nonfactorizable terms in the pole model. 
In Sec. III numerical results and discussions are presented.
A conclusion will be given in Sec. \ref{sec:con}.
Details on commutators in $S$-wave amplitudes are given in Appendix \ref{app:a}.   Baryon matrix elements and axial-vector form factors calculated in the MIT bag model are  summarized in Appendix \ref{app:b} and Appendix \ref{app:c},
respectively.

\section{Theoretical  framework}

\begin{figure}[t]
\begin{center}
\includegraphics[width=0.90\textwidth]{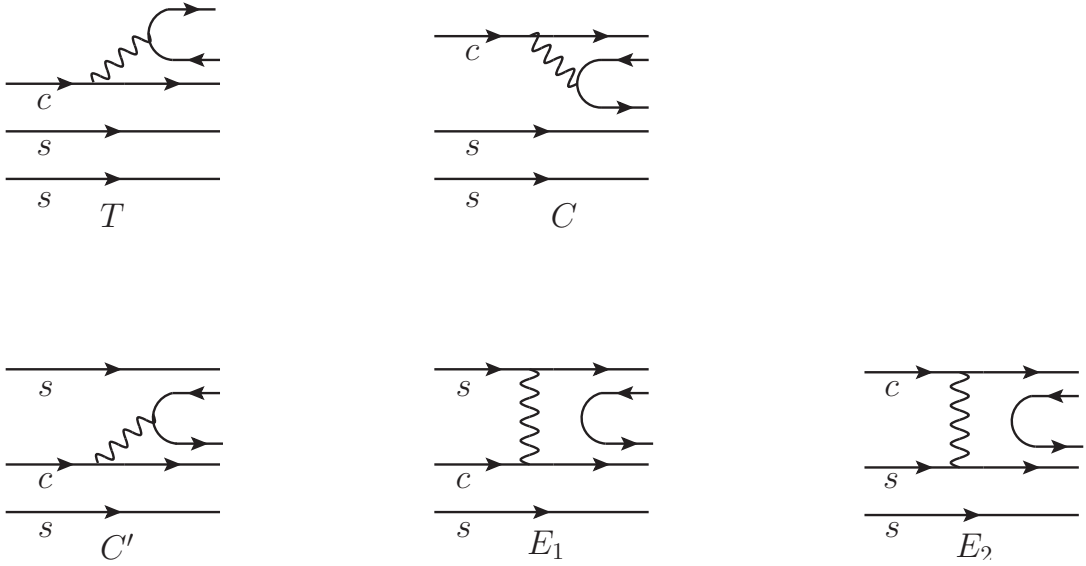}
\vspace{0.1cm}
\caption{Topological diagrams contributing to $\Omega_{c}\to \B +P$ decays: external $W$-emission $T$, internal $W$-emission $C$, inner $W$-emission $C'$,  $W$-exchange diagrams $E_1$ and $E_2$. }
 \label{fig:Omegac}
\end{center}
\end{figure}

In this work,
we  will continue studying weak decays of the $\Omega_c$ baryon
in the topological-diagram approach,
within which the factorizable and nonfactorizable contributions can be classified explicitly by topological diagrams \cite{Cheng:1991sn,Cheng:1993gf}.
Then different methods are adopted to calculate the two parts of contributions seperately.
The factorizable amplitudes are evaluated by naive factorization, while the pole model associated
with current algebra technique is applied in the calculation of nonfactorizable amplitudes.

\subsection{Topological diagrams}
 
The general formulation of the topological-diagram scheme for the nonleptonic weak decays of baryons was
proposed by Chau, Tseng and Cheng
more than two decades ago \cite{Chau:1995gk}, which was then applied to all the decays of the antitriplet and sextet charmed baryons.
Here we should emphasize that the topological diagram is not identical to the Feynman diagram, an example of which can be
found in \cite{Korner:1978tc}.  In the topological-diagram approach, even when 
final-state rescattering is included, we can still classify the diagrams 
according to their topology. 
In charmed meson decays, the extraction of the topological diagrams 
from the experimental data of Cabibbo-favored (CF) channels, together with SU(3) symmetry,  allows to predict  branching fractions of singly Cabibbo-suppressed (SCS)
and doubly Cabibbo-suppressed (DCS) decays and even  CP violation. For the charmed baryon decays, 
however, there are not adequate data on branching fractions
and decay asymmetries to extract the topological diagrams. Nevertheless, we can still use
 topological diagrams to identify factorizable and nonfactorizable decay amplitudes.

 For the weak decays $\Omega_{c}\to \mathcal{B}+P$ ($\mathcal{B}$ is baryon octet) of interest in this work, the relevant topological diagrams are
the external $W$-emission $T$, the internal $W$-emission $C$, the inner $W$-emission $C'$,  and the $W$-exchange diagrams $E_1$ as well as $E_2$ as depicted in Fig. \ref{fig:Omegac}. 
Among them, $T$ and $C$ are factorizable \footnote{
Strictly speaking, there are nonfactorizable contributions in the two diagrams.
However, in charm physics these nonfactorizable contributions  can be absorbed by  an effective $N$ in the effective Wilson coefficients $a_{1,2}=c_{1,2}+c_{2,1}/N $ ( see Eq. (9) as an example), and the value of $N$ can be extracted from the data. In that sense, the form of naive factorization can be kept 
 and hence $T$ and $C$ can be classified into factorizable ones.
}, 
while $C'$ and $W$-exchange give nonfactorizable contributions.  The relevant topological diagrams  for all   decay modes  of 
$\Omega_c$, 
including CF, SCS and DCS modes,
are shown  in Table \ref{tab:modes}.\footnote{
The reason for neglecting the so-called type-III diagram in \cite{Korner:1992wi}
or (d1), (d2) diagrams in \cite{Zenczykowski:1993jm} can be referred to \cite{Zou:2019kzq}.
}

We notice from Table \ref{tab:modes} that (i) among all the CF, SCS and DCS decays of $\Omega_c$, there is no  purely factorizable mode,
(ii) the modes containing $\Xi^{0}$ or $\Xi^-$ in final states  receive both factorizable and nonfactorizable contributions 
while the modes containing $\Sigma$ and $\Lambda^0$ in final states have purely nonfactorizable contribution, 
(iii) the $W$-exchange contribution 
is absent in the CF process but
manifests  in all the SCS and DCS decays,
and (iv) the two necleons $n$ and $p$, as parts of  the baryon octet, are absent in all the $\Omega_c$ decays.

\begin{table}[t]
\caption{Topological diagrams contributing to 
two-body weak decays
 $\Omega_c\to \mathcal{B} P$, where $\mathcal{B}$ is a baryon octet
and $P$ is a pseudoscalar meson.
%
\label{tab:modes}}
\vspace{0.3cm}
\begin{ruledtabular}
\begin{tabular}{l l l l l l}
\toprule
CF &  Contributions  & SCS & Contributions & DSC & Contributions \\
\colrule
$\Omega_c^0\to \Xi^0 \overline{K}^0$ & $C, C'$ &  $ \Omega_c^0\to \Xi^- \pi^+$  & $T,E_1$
& $ \Omega_{c}^0 \to \Xi^0 K^0 $ & $C, E_2$ \\
& 
& $\Omega_c^0\to \Sigma^+ K^-$  & $E_2$  & $\Omega_c^0 \to \Sigma^0\eta$
& $C', E_1, E_2$\\
$$ & $$ & $\Omega_c^0\to \Sigma^0 \overline{K}^0$ & $C',  E_2$ & $\Omega_c^0\to \Lambda^0 \eta$ & $C', E_1, E_2$\\
$$ & $$ & $\Omega_c^0\to \Lambda^0 \overline{K}^0$ & $C',  E_2$ & $\Omega_c^0\to \Sigma^0 \pi^0$ & $ E_1, E_2$\\
$$ & $$ & $\Omega_c^0\to \Xi^0 \pi^0$ & $C, E_1$ & $\Omega_c^0\to \Lambda^0 \pi^0$ & $ E_1, E_2$\\
$$ & $$ & 
& 
& $\Omega_c^0\to \Xi^- K^+$ & $T,  E_1$\\
$$ & $$ & $$ & $$ & $\Omega_c^0\to \Sigma^+ \pi^-$ & $E_2$\\
$$ & $$ & $$ & $$ & $\Omega_c^0\to \Sigma^- \pi^+$ & $ E_1$\\
\bottomrule
\end{tabular}
\end{ruledtabular}
\end{table}

\subsection{Kinematics}\label{sec:Kin}
The two-body weak decay $\mathcal{B}_i\to\mathcal{B}_f P$ 
generally can
be parametrized by both $S$- and $P$-amplitudes, giving
\begin{equation}
M(\mathcal{B}_i \to \mathcal{B}_f P)= i\ubar_f (A-B\gamma_5) u_i.
\end{equation}
Here $\mathcal{B}_i$ is actually $\Omega_c$,  the final state baryon $\mathcal{B}_f$ is a baryon octet,
and
 $P$ is a pseudoscalar meson.
The decay width and up-down decay asymmetry are given by
\begin{align}
& \Gamma=\frac{p_c}{8\pi}
\left[ \frac{(m_i+m_f)^2-m_P^2}{m_i^2}|A|^2
+\frac{(m_i-m_f)^2-m_P^2}{m_i^2}|B|^2\right], \nonumber \\
& \alpha=\frac{2\kappa {\rm{Re}} (A^* B)}{|A|^2+\kappa^2|B|^2},
\label{eq:Gamma}
\end{align}
where $p_c$ is the three-momentum
in the rest frame of the mother particle and $\kappa=p_c/(E_f+m_f)=\sqrt{(E_f-m_f)/(E_f+m_f)}$. The $S$- and $P$- wave amplitudes of the two-body decay generally receive both factorizable and non-factorizable contributions
\begin{align}
& A=A^{\rm{fac}}+A^{\rm{nf}},\quad
B=B^{\rm{fac}}+B^{\rm{nf}}.
\label{eq:amplitude}
\end{align}
We will discuss the details of factorizable and non-factorizable amplitudes  separately in below.

\subsection{Factorizable amplitudes}
Here the state-of-the-art effective Hamiltonian approach is adopted to describe factorizable contributions
in the charmed baryon decays.  Physical observables can be expressed as products 
of short-distance and long-distance parameters.
The short-distance physics is characterized by Wilson coefficients, 
which is obtained from integrating out high energy part of theory, and long-distance physics lies in the
hadronic matrix elements.

\subsubsection{General expression of factorizable amplitudes}

The effective Hamiltonian for CF process in $\Omega_c$ decay  is
\begin{equation}
\mathcal{H}_{\rm{eff}}=\frac{G_F}{\sqrt{2}}V_{cs}V_{ud}^*(c_1O_1+c_2O_2)+h.c.,
\label{eq:Hamiltonian}
\end{equation}
where the four-quark operators are given by
\begin{equation}
O_1=(\sbar c)(\ubar d),\quad
O_2=(\ubar c)(\sbar d),
\end{equation}
with $ (\qbar_1 q_2)\equiv \qbar_1\gamma_\mu(1-\gamma_5) q_2\nonumber$.
The
Wilson coefficients to the leading order are given by $c_1=1.346$ and $c_2=-0.636$ at $\mu=1.25\,\rm{GeV}$ and
$\Lambda_{\rm{MS}}^{(4)}=325\,{\rm{MeV}}$ \cite{Buchalla:1995vs}.
Under naive factorization
the amplitude can be written down as
\begin{equation}
M=\la \overline{K}^0\Xi^0|\mathcal{H}_{\rm{eff}}|\Omega_{c}\ra
=
\frac{G_F}{\sqrt{2}}V_{cs}V_{ud}^* a_{2} \la \overline{K}^0|(\sbar d)|0\ra \la \Xi^0|(\ubar c)|\mathcal{B}_{c}\ra, 
\label{eq:CF}
\end{equation}
where  $a_2=c_2+\frac{c_1}{N}$. 
The value of $N$ is taken as $N\approx 7$,
 following our previous study in \cite{Cheng:2018hwl}.
We also give the definition of $a_1$ as
$a_1=c_1+\frac{c_2}{N}$
for the purpose of describing
 SCS and DCS decays conveniently.
In terms of the decay constant
\begin{equation}
\la K (q)|\sbar\gamma_\mu(1-\gamma_5) d|0\ra = if_K q_\mu
\label{KFF}
\end{equation}
and baryon-baryon transition form factors defined by
\begin{eqnarray}
\la  \Xi^0(p_2)|\cbar\gamma_\mu(1-\gamma_5) u|\Omega_{c}(p_1)\ra
&=&\ubar_2 \left[ f_1(q^2) \gamma_\mu -f_2(q^2)i\sigma_{\mu\nu}\frac{q^\nu}{M}+f_3(q^2)\frac{q_\mu}{M}\right.\\
&&\hspace{0.5cm} -\left.\left(g_1(q^2)\gamma_\mu-g_2 (q^2)i\sigma_{\mu\nu}\frac{q^\nu}{M}+g_3(q^2)
\frac{q_\mu}{M}
\right)\gamma_5
\right]u_1,  \nonumber
\end{eqnarray}
with
the momentum transfer  $q=p_1-p_2$,
we obtain the factorizable amplitude as
\begin{equation}
M(\Omega_{c}\to \Xi^0\overline{K}^0)
=i\frac{G_F}{\sqrt{2}}a_{2} V_{ud}^*V_{cs} f_K \ubar_2(p_2)\left[(m_1-m_2)f_1(q^2)
+(m_1+m_2)g_1(q^2) \gamma_5\right]u_1(p_1).
\end{equation}
Contributions from the form factors $f_{3}$ and $g_{3}$ have been neglected with the  similar
reason given in footnote 1 of \cite{Cheng:2020wmk}.
More explicitly, the factorizable  $S$- and $P$-wave amplitudes read
\begin{align}
&A^{\rm{fac}}\big|_{\rm{CF}}=\frac{G_F}{\sqrt{2}}a_{2} V_{ud}^*V_{cs} f_K(m_{\Omega_{c}}-m_{\Xi^0}) f_1(q^2),  
\nonumber\\
&B^{\rm{fac}}\big|_{\rm{CF}}= -\frac{G_F}{\sqrt{2}}a_{2} V_{ud}^*V_{cs} f_K(m_{\Omega_{c}}+m_{\Xi^0})
g_1(q^2).
\end{align}

Likewise, the $S$- and $P$-wave amplitudes for SCS processes and DCS processes are given, respectively,  by
\begin{align} \label{eq:factSCS}
&A^{\rm{fac}}\big|_{\rm{SCS}}=\frac{G_F}{\sqrt{2}}a_{1,2} 
V_{uq}^*V_{cq} f_P(m_{\Omega_{c}}-m_{\mathcal{B}}) f_1(q^2),
\nonumber\\
&B^{\rm{fac}}\big|_{\rm{SCS}}= -\frac{G_F}{\sqrt{2}}a_{1,2} 
V_{uq}^*V_{cq} f_P(m_{\Omega_{c}}+m_{\mathcal{B}})
g_1(q^2),
\end{align}
and
\begin{align} \label{eq:factDCS}
&A^{\rm{fac}}\big|_{\rm{DCS}}=\frac{G_F}{\sqrt{2}}a_{1,2} 
V_{us}^*V_{cd} f_P(m_{\Omega_{c}}-m_{\mathcal{B}}) f_1(q^2),
\nonumber\\
&B^{\rm{fac}}\big|_{\rm{DCS}}= -\frac{G_F}{\sqrt{2}}a_{1,2} 
V_{us}^*V_{cd} f_P(m_{\Omega_{c}}+m_{\mathcal{B}})
g_1(q^2),
\end{align}
where the choice of $a_i$ depends on the meson in final states.
In the amplitude of SCS processes Eq. (\ref{eq:factSCS})  the flavor of the down-type quark $q$ ($q=d$ or $s$) depends on individual process. 

\subsubsection{Baryon transition form factors}

The baryon transition form factor (FF) parameter characterizes the transition between two baryons and is
calculated non-perturbatively. 
There have been some works on $\Omega_c$-${\cal B}$ transition FFs, including 
non-relativistic quark model (NRQM) \cite{PerezMarcial:1989yh}, 
heavy quark effective theory (HQET)\cite{Cheng:1995fe}, as well as light-front quark model (LFQM) \cite{Zhao:2018zcb}.
As we have emphasized that non-factorizable contribution gives significant contribution 
to the decay rates and decay asymmetries, 
 the relative sign between factorizable and non-factorizable amplitudes is critical.
 Hence an estimation of FFs as well as baryon matrix elements in a self-consistent convention is desirable. 
In this work we shall  work out both FFs and baryonic matrix elements 
in the framework of MIT bag model \cite{MIT}.

\begin{table}[t]
\begin{center}
\linespread{1.5}
\footnotesize{
 \caption{The calculated form factors in the MIT bag model at maximum four-momentum
 transfer squared $q^2=q^2_{\rm{max}}=(m_i-m_f)^2$ and $q^2=m_P^2$.
 } \label{tab:FF}
\vspace{0.3cm}
\begin{ruledtabular}
\begin{tabular}
{ ccccc|ccc}
\toprule
 modes & ($c\bar{q}$)&  $f_1(q_{\rm{max}}^2)$ &$ f_1(m_P^2)/f_1(q_{\rm{max}}^2)$ & $f_1(m_P^2)$ &  $g_1(q_{\rm{max}}^2)$ &
 $g_1(m_P^2)/g_1(q_{\rm{max}}^2)$ & $g_1(m_P^2)$  \\
\colrule
$\Omega_{c}^{0}\rightarrow\Xi^{0}   \overline{K}^{0}$ & $(c\bar{s})$  & $Y_1$&  $0.37$ & $0.32$  & $-\frac{1}{3}Y_2$  &   $0.54$ &$-0.14$  \\
$\Omega_{c}^{0}\rightarrow\Xi^{-} \pi^{+}$  & $(c\bar{d})$  & $Y_1$  &$0.29$ & $0.25$    & $-\frac{1}{3}Y_2$  &$0.46$  &$-0.12$ \\
$\Omega_{c}^{0}\rightarrow  \Xi^{0}   \pi^{0}$  & $(c\bar{d})$  & $Y_1$  &$0.28$& $0.25$   & $-\frac{1}{3}Y_2$  &$0.46$   &
$-0.12$\\
$\Omega_{c}^{0}\rightarrow  \Xi^{0} K^{0}$  & $(c\bar{d})$  & $Y_1$  &$0.32$  & $0.28$  & $-\frac{1}{3}Y_2$  &$0.50$ &$-0.13$  \\
$\Omega_{c}^{0}\rightarrow \Xi^{-}  K^{+}$  & $(c\bar{d})$  & $Y_1$  &$0.32$  & $0.28$ & $-\frac{1}{3}Y_2$  &$0.50$  & $-0.13$ \\
\bottomrule
\end{tabular}
\end{ruledtabular}
 }
\end{center}
\end{table}

Following \cite{Cheng:1991sn} we can write down the $q^2$ dependence of FF as
\begin{equation}
f_i(q^2)=\frac{f_i(0)}{(1-q^2/m_V^2)^2},\qquad
g_i(q^2)=\frac{g_i(0)}{(1-q^2/m_A^2)^2},
\label{eq:FF1}
\end{equation}
where $m_V
=2.01\,{\rm GeV}$, $m_A
=2.42\,{\rm GeV}$ for the  $(c\bar{d})$ quark content, and
$m_V=2.11\,{\rm GeV}$, $m_A=2.54\,{\rm GeV}$ for $(c\bar{s})$ quark content.
In the zero recoil limit where $q^2_{\rm{max}}=(m_i-m_f)^2$, FFs can be 
calculated in
the  MIT bag model \cite{Cheng:1993gf}, giving 
\begin{align} \label{eq:f1g1}
& f_1^{\mathcal{B}_f \mathcal{B}_i}(q^2_{\rm{max}})=\la \mathcal{B}_f^\uparrow| b_{q_1}^\dagger b_{q_2}| \mathcal{B}_i^\uparrow\ra \int d^3 \bm{r}\Big(u_{q_1}(r)u_{q_2}(r)+v_{q_1}(r)v_{q_2}(r)\Big),
\nonumber\\
& g_1^{\mathcal{B}_f \mathcal{B}_i}(q^2_{\rm{max}})=\la \mathcal{B}_f^\uparrow| b_{q_1}^\dagger b_{q_2} \sigma_z| \mathcal{B}_i^\uparrow\ra \int d^3 \bm{r}
\left(u_{q_1}(r)u_{q_2}(r)-\frac13v_{q_1}(r)v_{q_2}(r)\right),
\end{align}
where $u(r)$ and $v(r)$ are the large and small components, respectively, of the quark wave function in the bag model.
FFs at different $q^2$ are related via
\begin{equation}
f_i(q_2^2)=\frac{(1-q_1^2/m_V^2)^2}{(1-q_2^2/m_V^2)^2}f_i(q_1^2),\qquad
g_i(q_2^2)=\frac{(1-q_1^2/m_A^2)^2}{(1-q_2^2/m_A^2)^2}g_i(q_1^2).
\label{eq:FF2}
\end{equation}
This allows us to obtain the physical FFs  at $q^2=m_P^2$.

It is obvious that the FF at $q^2_{\rm{max}}$ is determined only by the baryons in initial and final states.
However, its evolution with $q^2$ is governed by both the final-state meson and relevant quark content. Such dependence is reflected in
Table \ref{tab:FF}, in which the quark contents are shown in the second column.
In the zero recoil limit, the FFs at $q^2_{\rm{max}}$ calculated from Eq. (\ref{eq:FF1})
are presented in the third and sixth columns. And then in the fourth and seventh columns, the  evolution of FFs from $q^2=q^2_{\rm{max}}$ to $q^2=m_P^2$ are derived according to Eq. (\ref{eq:FF2}).
The auxiliary quantities $Y_{1,2}$ can be obtained from the calculation in MIT bag model,  giving
\begin{align}
&
Y_1=4\pi \int r^2 dr (u_u u_c+v_u v_c),\quad~~ 
Y_1^s=4\pi \int  r^2 dr (u_s u_c+v_s v_c), \nonumber\\
&Y_2=4\pi \int r^2 dr (u_u u_c-\frac13 v_u v_c),\quad Y_2^s=4\pi \int  r^2  dr (u_s u_c-\frac13 v_s v_c).
\end{align}
The model parameters are adopted from \cite{Cheng:2018hwl} and references therein. Numerically, we have $Y_1=0.88, Y_1^s=0.95,
Y_2=0.77$
and $Y_2^s=0.86$, which are consistent with
the corresponding numbers in \cite{Cheng:1993gf}.

\begin{table}[t]
\caption{Form factors $f_1(q^2)$ and $g_1(q^2)$ at $q^2=0$ for  $\Omega_{c}^0\to \Xi$ transitions evaluated in the MIT bag model (this work),  non-relativistic quark model (NRQM) \cite{PerezMarcial:1989yh} and heavy quark effective theory (HQET)\cite{Cheng:1995fe}, which were quoted in \cite{Dhir:2015tja}, as well as light-front quark model (LFQM) \cite{Zhao:2018zcb}.
}
 \label{tab:FFcomp}
\begin{ruledtabular}
\begin{center}
{\footnotesize{
\begin{tabular}{l *4{c} |*4{c}}
\toprule
\multirow{2}{*}{${\Omega_c}\to {\cal B}$}    &
\multicolumn{4}{c}{$f_1(0)$ } &
\multicolumn{4}{c}{$g_1(0)$ } \\
\cline{2-5}\cline{5-9}
  & ~~MIT~~ & ~~NRQM~~ &~~HQET~~&~~LFQM~~   & ~~MIT~~ &~~NRQM~~  & ~~HQET~~  &~~LFQM~~ \\
\midrule
\colrule
$\Omega_{c}^{0}\to \Xi$     &  0.34  & 0.23 & 0.34 & 0.653
 & -0.15 & -0.14 & -0.10  &-0.182 \\
\bottomrule
\end{tabular}
}}
\end{center}
\end{ruledtabular}
\end{table}

A comparison is made  in Table \ref{tab:FFcomp}  among  results of  $\Omega_c\to \Xi$ FF in various approaches
at $q^2=0$.
From Eq. (\ref{eq:FF1}) and Table \ref{tab:FF}, FFs at $q^2=q^2_{\text{max}}$ are identical for $\Omega_c\to \Xi^{0,-} P$.
However, the values would be changed at different energy scale and also depend on different meson states.
In this comparison, we ignore such slight differences. 
Note that all the  nonperturbative quantities involving in this work are calculated in the MIT bag model, thus 
our convention of signs should be consistent globally.
For this reason we  correct the signs in both NRQM and HQET cases.
Apparently, NRQM gives small values  for both $f_1$ and $g_1$ , while the predictions from LFQM 
are the largest.

\subsection{Non-factorizable amplitudes}

Now it is widely accepted that in charmed baryon decays nonfactorizable amplitude can not be neglected, 
sometimes even gives dominated contributions to branching fraction and decay asymmetry in particular processes. In the topological-diagram approach, these nonfactorizable contributions have been presented by topological diagrams $C', E_1$ and $E_2$ in Fig. \ref{fig:Omegac}.  
A further calculation of nonfactorizable contribution 
relies on the pole model, while the pole diagrams will be obtained by the correspondence between topological diagrams \cite{Cheng:1993gf}.

The general formula for $S$- and $P$-wave non-factorizable amplitudes in the pole model is given by
\cite{Cheng:2018hwl,Zou:2019kzq,Cheng:2020wmk}
\begin{align}
& A^{\rm{pole}}=-\sum\limits_{B_n^*(1/2^-)}\left[\frac{g_{B_f B_n^* M}b_{n^* i}}{m_i-m_{n^*}} +
\frac{b_{fn^*}g_{B_{n}^*B_i M}}{m_f-m_{n^*}}\right],\nonumber\\
& B^{\rm{pole}}=\sum\limits_{B_n}\left[ \frac{g_{B_f B_n M}a_{ni}}{m_i-m_n}
+\frac{a_{fn}g_{B_n B_i M}}{m_f-m_n}\right],
\end{align}
with the baryonic matrix elements
\begin{equation}
\la \mathcal{B}_n|H|\mathcal{B}_i\ra = \ubar_n (a_{ni} + b_{n i}\gamma_5) u_i,\qquad
\la \mathcal{B}_i^*(1/2^-) | H|\mathcal{B}_j\ra =\ubar_{i*} b_{i^* j} u_j.
\end{equation}
To estimate  the $S$-wave amplitudes in the pole model is a difficult and nontrivial task as it involves the matrix elements and strong coupling constants of $1/2^-$ baryon resonances which is less known \cite{Cheng:1991sn}. 
Nevertheless, provided a soft emitted pseudoscalar meson, the intermediate excited baryons can be summed up, leading to a commutator term
\be
A^{\rm{com}} &=& -\frac{\sqrt{2}}{f_{P^a}}\la \mathcal{B}_f|[Q_5^a, H_{\rm{eff}}^{\rm PV}]|\mathcal{B}_i\ra
=\frac{\sqrt{2}}{f_{P^a}}\la \mathcal{B}_f|[Q^a, H_{\rm{eff}}^{\rm PC}]|\mathcal{B}_i\ra, \label{eq:Apole}
\en
with
\begin{equation}
Q^a=\int d^3x \qbar\gamma^0\frac{\lambda^a}{2}q,\qquad
Q^a_5=\int d^3x \qbar\gamma^0\gamma_5\frac{\lambda^a}{2}q.
\end{equation}
Likewise, the $P$-wave amplitude is reduced in the soft-meson limit to
\be
B^{\rm{ca}} &=& \frac{\sqrt{2}}{f_{P^a}}\sum_{\mathcal{B}_n}\left[ g^A_{\mathcal{B}_f \mathcal{B}_n}\frac{m_f+m_n}{m_i-m_n}a_{ni}
+a_{fn}\frac{m_i + m_n}{m_f-m_n} g_{\mathcal{B}_n \mathcal{B}_i}^A\right],
\label{eq:Bpole}
\en
where we have applied the generalized Goldberger-Treiman relation
\begin{equation} \label{eq:GT}
g_{_{\mathcal{B'B}P^a}}=\frac{\sqrt{2}}{f_{P^a}}(m_{\mathcal{B}}+m_{\mathcal{B'}})g^A_{\mathcal{B'B}}.
\end{equation}
Eqs. (\ref{eq:Apole}) and (\ref{eq:Bpole}) are the master equations for nonfactorizable amplitudes in the pole model under the soft meson approximation.

\subsubsection{$S$-wave amplitudes}

We have demonstrated that $S$-wave amplitudes can be simplified to baryon matrix elements 
of a set of commutators in the limit of soft meson, see Eq. (\ref{eq:Apole}). In Appendix \ref{app:a} more explicit expressions
for commutators corresponding to different final states are given.
The remaining task is then to evaluate different sets of commutators. 

We shall present our results after a straightforward calculation,
for the S-wave amplitudes, as follows:
\begin{align}
&{{A^{\rm{com}}(\Omega_{c}^{0}\rightarrow\Xi^{0}   \overline{K}^{0})= -\frac{\sqrt{2}}{f_K} a_{\Xi^{0}\Xi^{'0}_{c}}}},
\end{align}
for Cabibbo-favored (CF) process,
\begin{align}
&A^{\rm{com}}(\Omega_{c}^{0}\rightarrow\Xi^{-} \pi^{+})=  \frac{1}{f_\pi} a_{\Xi^0 \Omega_c^0},\hspace{3.6cm} A^{\rm{com}}(\Omega_{c}^{0}\rightarrow\Sigma^{+} K^{-})= \frac{1}{f_K} (a_{\Xi^0 \Omega_c^0}-\sqrt{2}a_{\Sigma^{+} \Xi_{c}^{'+}}),  \nonumber\\
&A^{\rm{com}}(\Omega_{c}^{0}\rightarrow\Sigma^{0} \overline{K}^{0})=\frac{1}{f_K} (-\frac{\sqrt{2}}{2} a_{\Xi^{0} \Omega_c^0}-\sqrt{2} a_{\Sigma^{0}\Xi^{'0}_c}), \quad
A^{\rm{com}}(\Omega_{c}^{0}\rightarrow \Xi^{0} \pi^{0})= \frac{1}{f_{\pi}} \frac{\sqrt{2}}{2}a_{\Xi^0 \Omega_c^0},
\nonumber\\
&A^{\rm{com}}(\Omega_{c}^{0}\rightarrow\Lambda^{0} \overline{K}^{0})=\frac{1}{f_K} (\frac{\sqrt{6}}{2} a_{\Xi^{0} \Omega_c^0}-\sqrt{2} a_{\Lambda^{0}\Xi^{'0}_c}),   
\end{align}
for singly Cabibbo-suppressed (SCS) procesees,
and
\begin{align}
&A^{\rm{com}}(\Omega_{c}^{0}\rightarrow  \Xi^{0} K^{0})=\frac{1}{f_K} (-\frac{\sqrt{2}}{2}a_{\Sigma^0 \Omega_c^0}+\frac{\sqrt{6}}{2}a_{\Lambda \Omega_c^0}), \quad
A^{\rm{com}}(\Omega_{c}^{0}\rightarrow\Sigma^{0} \eta)= \frac{ \sqrt{6}}{f_{\eta_{8}}}a_{\Sigma^0 \Omega_c^0} ,
\nonumber\\
&A^{\rm{com}}(\Omega_{c}^{0}\rightarrow\Lambda^{0} \eta)= \frac{ \sqrt{6}}{f_{\eta_{8}}}a_{\Lambda^0 \Omega_c^0} ,\hspace{3.7cm}
A^{\rm{com}}(\Omega_{c}^{0}\rightarrow\Sigma^{-} \pi^{+})= \frac{\sqrt{2}}{f_{\pi}}a_{\Sigma^0 \Omega_c^0} ,   \nonumber\\
&A^{\rm{com}}(\Omega_{c}^{0}\rightarrow\Xi^{-} K^{+})= -\frac{1}{f_{K}}(\frac{\sqrt{2}}{2}a_{\Sigma^0 \Omega_c^0}+\frac{\sqrt{6}}{2}a_{\Lambda \Omega_{c}^{0}}),\quad
A^{\rm{com}}(\Omega_{c}^{0}\rightarrow\Sigma^{+} \pi^{-})= -\frac{\sqrt{2}}{f_{\pi}}a_{\Sigma^0 \Omega_c^0} ,   \nonumber\\
&A^{\rm{com}}(\Omega_{c}^{0}\rightarrow\Sigma^{0} \pi^{0})= 0,\hspace{5cm}
A^{\rm{com}}(\Omega_{c}^{0}\rightarrow\Lambda^{0} \pi^{0})= 0,
\end{align}
for the doubly Cabibbo-suppressed (DCS) processes,
where the baryonic matrix element $\la \mathcal{B}'|H_{\rm{eff}}^{\rm PC}|\mathcal{B}\ra$ is denoted by $a_{\mathcal{B}'\mathcal{B}}$. We find that a straightforward calculation of the last two terms directly 
leads to vanishing results. This can be easily understood as $I_3(\Lambda^0)=I_3(\Sigma^0)=0$.


\subsubsection{P-wave amplitudes}

Now we turn to the nonfactorizable $P$-wave amplitudes given by
Eq. (\ref{eq:Bpole}). 
By substituting explicit hadron states,
we have
\begin{align}
&B^{\rm{ca}}(\Omega_{c}^{0}\rightarrow\Xi^{0}\overline{K}^{0})=\frac{1}{f_{K}}\left(a_{\Xi^{0}\Xi^{0}_{c}}\frac{m_{\Omega_{c}^{0}}+m_{\Xi_{c}^{0}}}{m_{\Xi^{0}}-m_{\Xi_{c}^{0}}}
g^{A(\overline{K}^{0})}_{\Xi_{c}^{0}\Omega_{c}^{0}}+a_{\Xi^{0}\Xi^{'0}_{c}}\frac{m_{\Omega_{c}^{0}}+m_{\Xi_{c}^{'0}}}{m_{\Xi^{0}}-m_{\Xi_{c}^{'0}}}
g^{A(\overline{K}^{0})}_{\Xi_{c}^{'0}\Omega_{c}^{0}}\right),
\end{align}
 for CF decays,
\begin{align}
&B^{\rm{ca}}(\Omega_{c}^{0}\rightarrow\Xi^{-}\pi^{+})=\frac{1}{f_{\pi}}\left(g^{A(\pi^{+})}_{\Xi^{-}\Xi^{0}}\frac{m_{\Xi^{-}}+m_{\Xi^{0}}}
{m_{\Omega^{0}_{c}}-m_{\Xi^{0}}}a_{\Xi^{0}\Omega^{0}_{c}}\right),\nonumber\\
&B^{\rm{ca}}(\Omega_{c}^{0}\rightarrow\Sigma^{+}K^{-})=\frac{1}{f_{K}}\left(g^{A(K^{-})}_{\Sigma^{+}\Xi^{0}}\frac{m_{\Sigma^{+}}+m_{\Xi^{0}}}
{m_{\Omega^{0}_{c}}-m_{\Xi^{0}}}a_{\Xi^{0}\Omega^{0}_{c}}\right),\nonumber\\
&B^{\rm{ca}}(\Omega_{c}^{0}\rightarrow\Sigma^{0}\overline{K}^{0})=\frac{1}{f_{K}}\left(a_{\Sigma^{0}\Xi^{0}_{c}}\frac{m_{\Omega_{c}^{0}}+m_{\Xi_{c}^{0}}}{m_{\Sigma^{0}}-m_{\Xi_{c}^{0}}}
g^{A(\overline{K}^{0})}_{\Xi_{c}^{0}\Omega_{c}^{0}}+a_{\Sigma^{0}\Xi^{'0}_{c}}\frac{m_{\Omega_{c}^{0}}+m_{\Xi_{c}^{'0}}}{m_{\Sigma^{0}}-m_{\Xi_{c}^{'0}}}
g^{A(\overline{K}^{0})}_{\Xi_{c}^{'0}\Omega_{c}^{0}}
+g^{A(\overline{K}^{0})}_{\Sigma^{0}\Xi^{0}}\frac{m_{\Sigma^{0}}+m_{\Xi^{0}}}{m_{\Omega_{c}^{0}}-m_{\Xi^{0}}}
a_{\Xi^{0}\Omega^{0}_{c}}\right),\nonumber\\
&B^{\rm{ca}}(\Omega_{c}^{0}\rightarrow\Lambda^{0}\overline{K}^{0})=\frac{1}{f_{K}}\left(a_{\Lambda^{0}\Xi^{0}_{c}}\frac{m_{\Omega_{c}^{0}}+m_{\Xi_{c}^{0}}}{m_{\Lambda^{0}}-m_{\Xi_{c}^{0}}}
g^{A(\overline{K}^{0})}_{\Xi_{c}^{0}\Omega_{c}^{0}}+a_{\Lambda^{0}\Xi^{'0}_{c}}\frac{m_{\Omega_{c}^{0}}+m_{\Xi_{c}^{'0}}}{m_{\Lambda^{0}}-m_{\Xi_{c}^{'0}}}
g^{A(\overline{K}^{0})}_{\Xi_{c}^{'0}\Omega_{c}^{0}}+g^{A(\overline{K}^{0})}_{\Lambda^{0}\Xi^{0}}\frac{m_{\Lambda^{0}}+m_{\Xi^{0}}}{m_{\Omega_{c}^{0}}-m_{\Xi^{0}}}
a_{\Xi^{0}\Omega^{0}_{c}}\right),\nonumber\\
&B^{\rm{ca}}(\Omega_{c}^{0}\rightarrow\Xi^{0}\pi^{0})=\frac{\sqrt{2}}{f_{\pi}}\left(g^{A(\pi^{0})}_{\Xi^{0}\Xi^{0}}\frac{m_{\Xi^{0}}+m_{\Xi^{0}}}
{m_{\Omega^{0}_{c}}-m_{\Xi^{0}}}a_{\Xi^{0}\Omega^{0}_{c}}\right),
\end{align}
for SCS processes, and
\begin{align}
&B^{\rm{ca}}(\Omega_{c}^{0}\rightarrow\Xi^{0}K^{0})=\frac{1}{f_{K}}\left(g^{A(K^{0})}_{\Xi^{0}\Lambda^{0}}\frac{m_{\Xi^{0}}+m_{\Lambda^{0}}}
{m_{\Omega^{0}_{c}}-m_{\Lambda^{0}}}a_{\Lambda^{0}\Omega^{0}_{c}}+g^{A(K^{0})}_{\Xi^{0}\Sigma^{0}}\frac{m_{\Xi^{0}}+m_{\Sigma^{0}}}
{m_{\Omega^{0}_{c}}-m_{\Sigma^{0}}}a_{\Sigma^{0}\Omega^{0}_{c}}\right),\nonumber\\
&B^{\rm{ca}}(\Omega_{c}^{0}\rightarrow\Sigma^{0}\eta_{8})=\frac{\sqrt{2}}{f_{\eta_{8}}}\left(a_{\Sigma^{0}\Omega_{c}^{0}}\frac{m_{\Omega^{0}_{c}}+m_{\Omega^{0}_{c}}}{m_{\Sigma^{0}}-
m_{\Omega_{c}^{0}}}
g_{\Omega_{c}^{0}\Omega_{c}^{0}}^{A(\eta_{8})}+g^{A(\eta_{8})}_{\Sigma^{0}\Sigma^{0}}\frac{m_{\Sigma^{0}}+m_{\Sigma^{0}}}
{m_{\Omega^{0}_{c}}-m_{\Sigma^{0}}}a_{\Sigma^{0}\Omega^{0}_{c}}+g^{A(\eta_{8})}_{\Sigma^{0}\Lambda^{0}}\frac{m_{\Sigma^{0}}+m_{\Lambda^{0}}}
{m_{\Omega^{0}_{c}}-m_{\Lambda^{0}}}a_{\Lambda^{0}\Omega^{0}_{c}}\right), \nonumber\\
&B^{\rm{ca}}(\Omega_{c}^{0}\rightarrow\Lambda^{0}\eta_{8})=\frac{\sqrt{2}}{f_{\eta_{8}}}\left(a_{\Lambda^{0}\Omega_{c}^{0}}\frac{m_{\Omega^{0}_{c}}+
m_{\Omega^{0}_{c}}}{m_{\Lambda^{0}}-m_{\Omega_{c}^{0}}}g_{\Omega_{c}^{0}\Omega_{c}^{0}}^{A(\eta_{8})}+g^{A(\eta_{8})}_{\Lambda^{0}\Sigma^{0}}\frac{m_{\Lambda^{0}}+m_{\Sigma^{0}}}
{m_{\Omega^{0}_{c}}-m_{\Sigma^{0}}}a_{\Sigma^{0}\Omega^{0}_{c}}+g^{A(\eta_{8})}_{\Lambda^{0}\Lambda^{0}}\frac{m_{\Lambda^{0}}+m_{\Lambda^{0}}}
{m_{\Omega^{0}_{c}}-m_{\Lambda^{0}}}a_{\Lambda^{0}\Omega^{0}_{c}}\right), \nonumber\\
&B^{\rm{ca}}(\Omega_{c}^{0}\rightarrow\Xi^{-}K^{+})=\frac{1}{f_{K}}\left(g^{A(K^{+})}_{\Xi^{-}\Sigma^{0}}\frac{m_{\Xi^{-}}+m_{\Sigma^{0}}}
{m_{\Omega^{0}_{c}}-m_{\Sigma^{0}}}a_{\Sigma^{0}\Omega^{0}_{c}}+g^{A(K^{+})}_{\Xi^{-}\Lambda^{0}}\frac{m_{\Xi^{-}}+m_{\Lambda^{0}}}
{m_{\Omega^{0}_{c}}-m_{\Lambda^{0}}}a_{\Lambda^{0}\Omega^{0}_{c}}\right),\nonumber\\
&B^{\rm{ca}}(\Omega_{c}^{0}\rightarrow\Sigma^{-}\pi^{+})=\frac{1}{f_{\pi}}\left(g^{A(\pi^{+})}_{\Sigma^{-}\Sigma^{0}}\frac{m_{\Sigma^{-}}+m_{\Sigma^{0}}}
{m_{\Omega^{0}_{c}}-m_{\Sigma^{0}}}a_{\Sigma^{0}\Omega^{0}_{c}}+g^{A(\pi^{+})}_{\Sigma^{-}\Lambda^{0}}\frac{m_{\Sigma^{-}}+m_{\Lambda^{0}}}
{m_{\Omega^{0}_{c}}-m_{\Lambda^{0}}}a_{\Lambda^{0}\Omega^{0}_{c}}\right),\nonumber\\
&B^{\rm{ca}}(\Omega_{c}^{0}\rightarrow\Sigma^{+}\pi^{-})=\frac{1}{f_{\pi}}\left(g^{A(\pi^{-})}_{\Sigma^{+}\Sigma^{0}}\frac{m_{\Sigma^{+}}+m_{\Sigma^{0}}}
{m_{\Omega^{0}_{c}}-m_{\Sigma^{0}}}a_{\Sigma^{0}\Omega^{0}_{c}}+g^{A(\pi^{-})}_{\Sigma^{+}\Lambda^{0}}\frac{m_{\Sigma^{+}}+m_{\Lambda^{0}}}
{m_{\Omega^{0}_{c}}-m_{\Lambda^{0}}}a_{\Lambda^{0}\Omega^{0}_{c}}\right),\nonumber\\
&B^{\rm{ca}}(\Omega_{c}^{0}\rightarrow\Sigma^{0}\pi^{0})=\frac{\sqrt{2}}{f_{\pi}}\left(g^{A(\pi^{0})}_{\Sigma^{0}\Lambda^{0}}\frac{m_{\Sigma^{0}}+m_{\Lambda^{0}}}
{m_{\Omega^{0}_{c}}-m_{\Lambda^{0}}}a_{\Lambda^{0}\Omega^{0}_{c}}+g^{A(\pi^{0})}_{\Sigma^{0}\Sigma^{0}}\frac{m_{\Sigma^{0}}+m_{\Sigma^{0}}}
{m_{\Omega^{0}_{c}}-m_{\Sigma^{0}}}a_{\Sigma^{0}\Omega^{0}_{c}}\right),\nonumber\\
&B^{\rm{ca}}(\Omega_{c}^{0}\rightarrow\Lambda^{0}\pi^{0})=\frac{\sqrt{2}}{f_{\pi}}\left(g^{A(\pi^{0})}_{\Lambda^{0}\Lambda^{0}}\frac{m_{\Lambda^{0}}+m_{\Lambda^{0}}}
{m_{\Omega^{0}_{c}}-m_{\Lambda^{0}}}a_{\Lambda^{0}\Omega^{0}_{c}}+g^{A(\pi^{0})}_{\Lambda^{0}\Sigma^{0}}\frac{m_{\Lambda^{0}}+m_{\Sigma^{0}}}
{m_{\Omega^{0}_{c}}-m_{\Sigma^{0}}}a_{\Sigma^{0}\Omega^{0}_{c}}\right),
\end{align}
for DCS decay processes.
A furthermore derivation of non-perturbative quantities 
$a_{\mathcal{B}\mathcal{B}'}$ and $g_{\B \B'}^A$ can be found in Appendices \ref{app:b} and \ref{app:c}.

\section{Results and discussion}

\subsection{Numerical results and discussions}

In this section, we shall numerically calculate branching fractions and up-down decay asymmetries. 
The decay asymmetries rely on $S$- and $P$-wave amplitudes, which have been calculated analytically yet. One more parameter, lifetime, enters the calculation of branching fractions based on the decay width in Eq. (\ref{eq:Gamma}).
The value of the lifetime quoted in this work is reported by LHCb in 2018 (see Eq. (\ref{eq:lifetime})).

\begin{table}[b]
 \caption{
 Decays $\Omega_{c}\to \mathcal{B} P$:  the amplitudes are in units of  $10^{-2}G_F  {\rm{GeV}}^2$,
branching fractions for CF(SCS, DCS) process(es) is (are) in unit(s) of $10^{-2}$($10^{-3}$, $10^{-4}$), and the asymmetry parameters $\alpha$ are shown in the last column. \footnote{The two DCS channels $\Omega_c\to \Sigma^0\eta$ and $\Lambda^0 \pi^0$ are not included in the table for their vanishing $S$-wave, tiny $P$-wave
amplitudes, and hence almost zero branching fractions.}
 } \label{tab:result1}
 \vspace{0.3cm}
\begin{ruledtabular}
\begin{tabular}
{ l | c c c c c c | c c |c }
 \toprule
 Channel & $A^{\rm{fac}}$ &  $A^{\rm{com}}$ & $A^{\rm{tot}}$ & $B^{\rm{fac}}$ &  $B^{\rm{ca}}$ & $B^{\rm{tot}}$ & $\mathcal{B}_{\rm{theo}}$
 & $\mathcal{B}_{\rm{expt}}$ &  $\alpha_{\rm{theo}}$ \\
  \colrule
$\Omega_{c}^{0}\to\Xi^{0}\overline{K}^{0}$ & $ -2.15$  & $10.92$ & $8.78$ & $-2.64$ & $10.12$ & $7.48$ & $3.78$  & $-$  & $0.51$
\\
\hline
$\Omega_{c}^{0}\to\Sigma^{+}K^{-}$ & $0$  & $-0.01$ & $-0.01$ & $0$ & $-6.10$ & $-6.10$ & $2.32$  & $-$  & $0.01$
\\
$\Omega_{c}^{0}\to\Sigma^{0}\overline{K}^{0}$ & $0$  & $0.01$ & $0.01$ & $0$ & $-1.21$ & $-1.21$ & $0.09$  & $-$  & $-0.03$
\\
$\Omega_{c}^{0}\to\Lambda^{0}\overline{K}^{0}$ & $0$  & $-4.21$ & $-4.21$ & $0$ & $0.04$ & $0.04$ & $8.05$  & $-$  & $-0.01$
\\
$\Omega_{c}^{0}\to\Xi^{0}\pi^{0}$ & $-0.88$  & $-2.43$ & $-3.31$ & $-1.21$ & $1.00$ & $-0.21$ & $5.46$  & $-$  & $0.04$
\\
$\Omega_{c}^{0}\to\Xi^{-}\pi^{+}$ & $ -0.89$  & $-3.44$ & $-4.33$ & $-1.22$ & $1.42$ & $0.20$ & $9.34$  & $-$  & $-0.03$
\\
\hline
$\Omega_{c}^{0}\to\Xi^{-}K^{+}$ & $0.10$  & $1.34$ & $1.43$ & $0.13$ & $0.49$ & $0.62$ & $9.58$  & $-$  & $0.27$
\\
$\Omega_{c}^{0}\to\Xi^{0}K^{0}$ & $0.10$  & $-1.34$ & $-1.24$ & $0.131$ & $-0.49$ & $-0.36$ & $7.04 $  & $-$  & $0.18$
\\
$\Omega_{c}^{0}\to\Lambda \eta$ & $0$  & $-2.66$ & $-2.66$ & $0$ & $-2.56$ & $-2.56$ &$36.28$    & $-$  & $0.66$
\\
$\Omega_{c}^{0}\to\Sigma^{0} \pi^{0}$ & $0$  & $0$ & $0$ & $0$ & $-1.03$ & $-1.03$ & $0.77 $  & $-$  & $0$
\\
$\Omega_{c}^{0}\to\Sigma^{+} \pi^{-}$ & $0$  & $0$ & $0$ & $0$ & $-1.03$ & $-1.03$ & $0.77$  & $-$  & $0$
\\
$\Omega_{c}^{0}\to\Sigma^{-} \pi^{+}$ & $0$  & $0$ & $0$ & $0$ & $-1.03$ & $-1.03$ & $0.77$  & $-$  & $0$
\\
\bottomrule
\end{tabular}
\end{ruledtabular}
\end{table}

Factorizable  and nonfactorizable amplitudes,
 branching fractions and decay asymmetries of all the two-body weak decays of $\Omega_c$ , including CF, SCS and 
DCS processes,  are summarized in Table \ref{tab:result1}. 
The channel $\Omega_c\to \Xi^0\overline{K}^0$ is the unique CF mode among all the $\Omega_{c}\to \mathcal{B} P$
decays, where $\mathcal{B}$ is a baryon octet. In both $S$- and $P$- wave amplitudes, the nonfactorizable contributions are large and give destructive interference between factorizable ones.  The branching fraction with 
full factorizable and nonfactorizable contributions is predicted to be $3.78\%$.  The benchmark channel   
$\Omega_c\to\Omega^-\pi^+$, which 
is also classified into CF modes, proceeds through external $W$-emission and hence receives only factorizable contribution.  Naively, if the nonfactorizable terms of $\Omega_c\to \Xi^0\overline{K}^0$
are turned off, the predicted value of its branching fraction would be $2.44\%$.
This partially helps to understand the Belle measurement of large relative ratio between 
$\Omega_c\to\Xi^0\overline{K}^0$ and
 $\Omega_c\to\Omega^-\pi^+$,  see  Eq. (\ref{eq:ratio}).
 A detailed consideration for these channels decaying into baryon decuplet will be presented in a separate work.
 \footnote{
 A prediction for branching fraction of $\Omega_c\to\Omega^- \pi^+$,
 based on an early work in \cite{Xu:1992sw} with updated $a_1=1.26$ and latest $\Omega_c$
 lifetime, is of order $9\%$.
 The incompatibility among our prediction of $\Omega_c\to \Xi^0\overline{K}^0$, the prediction of
 $\Omega_c\to\Omega^- \pi^+$ in \cite{Xu:1992sw} and Belle measurement Eq. (\ref{eq:ratio}) will also
 be discussed therein.
  }
Although no explicit measurement of the mode $\Omega_c\to\Xi^0\overline{K}^0$ has been given,
a large branching fraction prediction indicates a direct measurement is promising in the near future. 
The decay asymmetry $\alpha$ is predicted to be positive and with a measurable value $0.51$, which is also testable when 
more data are available. 

The three decay modes $\Omega_c\to\Sigma^+ K^-, \Sigma^0\overline{K}^0, \Lambda^0\overline{K}^0$ in SCS channels, which do not receive factorizable contributions,
are typical examples for the essential role  of nonfactorizable contribution  in charmed baryon decays. 
 Due to the breaking SU(3) flavor symmetry, see parameter  $X_1^s$, the $S$-wave amplitudes for the modes with $\Sigma$ baryon final states are tiny but not vanishing. On the other hand, $S$-wave amplitude for  $\Omega_c\to\Lambda^0\overline{K}^0$ is significantly enhanced from $\Sigma^0 \overline{K}^0$ for its typical size is described by $X_2^s$, which is two orders of magnitude larger than $X_1^s$.
 Among the $P$-wave terms of the three modes, $\Sigma^+ K^-$ is the largest one as 
cancellation occurs in the other two modes.
Since $S$-wave amplitude dominates branching fraction, according to Eq. (\ref{eq:Gamma}),  
$\Omega_c\to\Lambda^0\overline{K}^0$ is predicted with the largest branching fraction among the three modes.
However, the decay asymmetries for all the three modes is tiny, which is a natural consequence of tiny value for either  $A$ or $B$.  The remaining two modes in SCS processes, $\Omega_c\to\Xi^0 \pi^0$ and $\Xi^- \pi^+$, 
share almost same factorizable  contributions while in both channels
the nonfactorizable terms contribute constructively in $S$-wave and destructively in $P$-wave terms,
leading to tiny decay asymmetries again.

Predictions for DCS channels are also summarized in Table \ref{tab:result1}. 
The mode $\Omega_c\to\Lambda^0\eta$ is of particular interest 
in all the DCS channels.
Its $S$- and $P$-wave amplitudes, which both are depicted by $X_2^D$, are substantial and hence 
lead to  a large branching fraction $0.36\%$. The large and positive decay asymmetry 0.66 is also predicted.
The $S$-wave amplitudes for the channels 
$\Omega_c\to \Sigma \pi/\eta $ and $\Omega_c\to \Lambda \pi^0$ vanish 
among all the modes which contain net nonfactorizable contribution due to
the reasons $I_3(\Sigma^0)=I_3(\Lambda)=0$ or $X_1^D=0$. Then the two decays
$\Omega_c\to\Sigma^0\eta $ and $\Lambda^0\pi^0$
are prohibited due to the 
properties $g^{A(\pi^0)}_{\Lambda^0\Lambda^0}= g^{A(\eta_8)}_{\Sigma^0\Lambda^0}=0$
furthermore.
The identical predictions for the three $\Sigma \pi$  modes are natural consequences in the pole model.
The vanishing $S$-wave amplitudes of the three modes leads to their null decay asymmetries,
while the  identical branching fractions are caused by their similar pole diagrams associated with isospin factor $1/\sqrt{2}$.

\subsection{Comparison with other works}

In the early 1990s, there were many efforts to study charmed baryon decays, among which 
few   were $\Omega_c$ involved \cite{Cheng:1993gf,Korner:1992wi,Ivanov:1997ra,Xu:1992sw}. 
Later semileptonic decays of heavy $\Omega$ baryons, including $\Omega_c$, was studied in \cite{Zhao:2018zcb, Pervin:2006ie}.
In recent years  there have been some interests on its  hadronic weak decays \cite{Dhir:2015tja,Gutsche:2018utw,Zhao:2018zcb}, in which \cite{Dhir:2015tja} focused on modes with
axial-vector final state and only modes with decuplet baryon final state were partially involved in \cite{Gutsche:2018utw}.
In Table \ref{tab:comparison}  a comparison with other groups, whose predictions have been updated by incorporating current $\Omega_c$ lifetime,  is summarized in available channels.

The CF channel  $\Omega_c\to\Xi^0\overline{K}^0$ attracts more attentions in the past \cite{Cheng:1993gf,Korner:1992wi,Ivanov:1997ra} and nonfactorizable contributions have been incorporated by all groups.
Based on the pole mode combing current algebra, our results both for branching fraction and decay asymmetry can 
be confirmed by the early calculation within the same approach \cite{Cheng:1993gf}.
However, the  prediction for branching fraction is around 10 times larger than an early estimation relied on pure pole model \cite{Cheng:1993gf}, and the sign of decay asymmetry is opposite. 
Small branching fraction and negative asymmetry were also predicted within a relativistic three-quark model with a Gaussian shape for the momentum dependence of baryon-three-quark vertex \cite{Ivanov:1997ra}. 
Such situation  occurred in the studies on anti-triplet charmed baryons
\cite{Cheng:2018hwl, Zou:2019kzq}. Taking the mode $\Lambda_c\to\Sigma^+\pi^0$ as an example, the  pure pole model  in \cite{Cheng:1993gf} and quark model calculation in \cite{Ivanov:1997ra} both predicted positive decay asymmetry
while current algebra predicted $\alpha=-0.76$, which
is consistent with experimental value
 $\alpha=-0.55\pm0.11$  \cite{Tanabashi:2018oca}.
Interestingly, working in an independent approach, 
K\"orner-Kr\"amer gave a consistent prediction  for 
both branching fraction and decay asymmetry in covariant quark model \cite{Korner:1992wi}.
 
 The branching fractions of SCS process $\Omega_c\to \Xi^-\pi^+$  and DCS process $\Omega_c\to \Xi^- K^+$ were estimated in \cite{Zhao:2018zcb}, where  baryon-baryon transition form factors were calculated in light-front quark model and
  only factorizable contribution 
was taken into account.  It has been widely accepted that nonfactorizable contribution should play an essential role in the
hadronic decays. The numerical results for each individual term in Table \ref{tab:result1} shows that nonfactorizable terms even give dominated contributions, which helps to explain why our prediction is more than 10 times larger.

\begin{table}[t]
\begin{center}
\caption{Predicted branching fractions in the unit of $10^{-2}, 10^{-3}$ and $10^{-4}$ (upper entry in each mode) and decay asymmetry $\alpha$ (lower entry) of  $\Omega_c$ decays by different groups.
}\label{tab:comparison}
 \vspace{0.3cm}
\begin{ruledtabular}
\begin{tabular}{ l c c c c c c }
\toprule
  Mode &  Our  & Cheng \textit{et al.}  & Cheng \textit{et al.} & K\"orner \textit{et al.} & Ivanov \textit{et al.}  & Zhao  \\
  & & CA \cite{Cheng:1993gf} & pole model \cite{Cheng:1993gf} &
   \cite{Korner:1992wi} & \cite{Ivanov:1997ra} & \cite{Zhao:2018zcb}  \\ 
\colrule
$\Omega_{c}\to \Xi^0 \overline{K}^0$  & 3.78 & 2.63 &0.35 & 4.69  & 0.09 & \\
& 0.51  & 0.44 & -0.93  & 0.51 &-0.81 & \\
$\Omega_{c}\to \Xi^- \pi^+$  & 9.34  &  &  &   &   & 0.7\\
& -0.03   &  & &  &  \\
$\Omega_{c}\to \Xi^- K^+$  & 9.58 &  &  &   &   & 0.6 \\
&  0.27 &  & &  &  \\
\bottomrule
\end{tabular}
\end{ruledtabular}
\end{center}
\end{table}

\section{Conclusions}\label{sec:con}

In this work we have systematically studied the branching fractions and up-down decay asymmetries of CF, SCS
and DCS decays of $\Omega_c$, the heaviest singly charmed baryon which decays weakly.
Both factorizable and nonfactorizable terms 
have been taken into account in the calculation of
$S$- and $P$-wave amplitudes.
To estimate nonfactorizable contribution, we work in the pole model for $P$-wave amplitudes and current algebra 
for $S$-wave ones. All the non-perturbative parameters, including baryon-baryon transition form factors, baryon matrix elements and axial-vector form factors, are evaluated within MIT bag model throughout the whole calculations. 

Some conclusions can be drawn from our analysis as follows. 
\begin{itemize}
\item 
The channel $\Omega_c\to \Xi^0\overline{K}^0$ is the unique mode for CF decay. 
Although no absolute branching fraction has been measured up to now, 
the predicted large value for branching fraction 
indicates this mode is quite promising to be measured in the near future. Meanwhile its 
decay asymmetry is predicted to be large in magnitude and positive in sign. 

\item Among all the SCS modes, the channel $\Omega_c\to \Lambda^0\overline{K}^0$ 
is special as it proceeds only through the nonfactorizable contributions.
Though $P$-wave amplitude is small, its  large $S$-wave amplitude leads to a large branching ratio.
Hence a measurement of $\Omega_c\to \Lambda^0\overline{K}^0$ in the future will demonstrate
the essential role of nonfactorizable contribution in charmed baryon weak decays.

\item The decay asymmetries of all SCS modes are small, which is a natural consequence of 
either small  $S$- or $P$-wave amplitude. In other words, it is difficult to measure decay asymmetries 
of  SCS $\Omega_c$ weak decay in experiment. 

\item 
The measurement of $\Omega_c\to \Lambda \eta$ will also be interesting.
Although this mode is classified as the DCS mode, not only the branching fraction is predicted to be large,
but also its decay asymmetry is predicted to be large in magnitude  and positive in sign, which makes the 
measurement  conceivable in experiment.
Such features can be explained by simultaneous large $S$- and $P$-wave amplitudes.

\item 
The two DCS modes $\Omega_c\to \Sigma^0\eta$ and $\Omega_c\to \Lambda^0\pi^0$  are forbidden, for
both $S$- and $P$-wave amplitudes are found to be zero. On the other hand, these two modes 
can serve as golden channels for new physics searching.

\end{itemize}

\begin{acknowledgments}

We would like to thank Prof. Hai-Yang Cheng for his encouragement and fruitful discussion of this work. 
This research  is supported by NSFC under Grant No. U1932104.

\end{acknowledgments}

\appendix

\section{Commutators in $S$-wave amplitudes}
\label{app:a}
The nonfactorizable $S$-wave amplitude is determined by the commutator terms of conserving charge $Q^a$ and the parity-conserving part of the Hamiltonian, shown in Eq. (\ref{eq:Apole}).
In terms of such commutators, we further present the $A^{\rm com}$  for various meson production more explicitly:
\begin{align}
&A^{\rm{com}}(B_i\to B_f \pi^{\pm})=\frac{1}{f_\pi}\la B_f|[I_{\mp}, H_{\rm{eff}}^{PC}]|B_i\ra,\\
&A^{\rm{com}}(B_i\to B_f \pi^{0})=\frac{\sqrt{2}}{f_\pi}\la B_f|[I_3, H_{\rm{eff}}^{PC}]|B_i\ra,\\
&A^{\rm{com}}(B_i\to B_f \eta_8)=\sqrt{\frac32}\frac{1}{f_{\eta_8}}\la B_f|[Y, H_{\rm{eff}}^{PC}]|B_i\ra, \label{eq:commu}\\
&A^{\rm{com}}(B_i\to B_f K^{\pm})=\frac{1}{f_K}\la B_f|[V_{\mp}, H_{\rm{eff}}^{PC}]|B_i\ra,\\
&A^{\rm{com}}(B_i\to B_f \overline{K}^0 )=\frac{1}{f_K}\la B_f|[U_{+}, H_{\rm{eff}}^{PC}]|B_i\ra,\\
&A^{\rm{com}}(B_i\to B_f {K^0})=\frac{1}{f_K}\la B_f|[U_{-}, H_{\rm{eff}}^{PC}]|B_i\ra.
\end{align}
In Eq. (\ref{eq:commu}),  $\eta_8$ is the octet component of the $\eta$ and $\eta'$
\be
\eta=\cos\theta\eta_8-\sin\theta\eta_0, \qquad \eta'=\sin\theta\eta_8+\cos\theta\eta_0,
\en
with $\theta=-15.4^\circ$ \cite{Kroll}. For the decay constant $f_{\eta_8}$,  we shall follow \cite{Kroll} to use $f_{\eta_8}=f_8\cos\theta$ with $f_8=1.26 f_\pi$.
The convention for hypercharge $Y$ is taken to be $Y=B+S-C$ \cite{Cheng:2018hwl}.

\section{Hadronic matrix elements}
\label{app:b}

The  baryonic matrix elements $a_{\B'\B}$ get involved  both in $S$-wave
and $P$-wave amplitudes. 
Their general expression in terms of the effective Hamiltonian Eq. (\ref{eq:Hamiltonian}) is given by
\begin{align}
&a_{\B'\B} \equiv  \la \B'|\mathcal{H}_{\rm{eff}}^{\rm{PC}}|\B\ra
=\left\{\begin{array}{ll}
\frac{G_F}{2\sqrt{2}} V_{cs} V^*_{ud} c_-\la \B' |O_- |\B\ra,& \text{CF} \\ \\
\frac{G_F}{2\sqrt{2}}\sum\limits_q V_{cq} V^*_{uq} c_-\la \B' |O^q_- |\B\ra,& \text{SCS} \\  \\
\frac{G_F}{2\sqrt{2}} V_{cd} V^*_{us} c_-\la \B' |O^D_- |\B\ra,& \text{DCS}
\end{array}\right.
\end{align}
where $O_\pm=(\bar{s}c)(\bar{u}d)\pm(\bar{s}d)(\bar{u}c)$,
$O^q_\pm=(\bar{q}c)(\bar{u}q)\pm(\bar{q}q)(\bar{u}c)$  (with $q=d, s$) and
$O^D_\pm=(\bar{d}c)(\bar{u}s)\pm(\bar{d}s)(\bar{u}c)$
 and $c_\pm=c_1\pm c_2$.
The matrix element of $O^{(q,D)}_+$ vanishes as this operator is symmetric in color indices. 
We shall calculate relevant baryon matrix elements in MIT bag model.
The results for CF processes are
\begin{equation}
  \langle\Xi^{0}|O_{-}|\Xi^{'0}_{c}\rangle=-\frac{2\sqrt{2}}{3}(X_{1}+9X_{2})(4\pi),
  \qquad
   \langle\Xi^{0}|O_{-}|\Xi^{0}_{c}\rangle=\frac{2\sqrt{6}}{3}(X_{1}-3X_{2})(4\pi),
\end{equation}
Likewise, the matrix elements for SCS decays are calculated to be
\begin{align}
 & \langle\Xi^{0}|O_{-}^{d}|\Omega_{c}^{0}\rangle=0,\hspace{3cm}
  \langle\Xi^{0}|O_{-}^{s}|\Omega_{c}^{0}\rangle=-\frac{4}{3}(X_{1}^{s}+9X_{2}^{s})(4\pi),\nonumber\\
 & \langle\Sigma^{+}|O_{-}^{d}| \Xi_{c}^{'+}\rangle=0, \hspace{2.7cm}
  \langle\Sigma^{+}|O_{-}^{s}| \Xi_{c}^{'+}\rangle=\frac{2\sqrt{2}}{3}(X_{1}^{s}-9X_{2}^{s})(4\pi),\nonumber\\
 & \langle\Sigma^{0}|O_{-}^{d}| \Xi^{'0}_{c}\rangle=\frac{4}{3}X_{1}^{d}(4\pi),\hspace{1.6cm}
  \langle\Sigma^{0}|O_{-}^{s}| \Xi^{'0}_{c}\rangle=-\frac{2}{3}(X_{1}^{s}-9X_{2}^{s})(4\pi),\nonumber\\
 &\langle\Lambda^{0}|O_{-}^{d}| \Xi^{'0}_{c}\rangle=-4\sqrt{3}X_{2}^{d}(4\pi),\hspace{1.0cm}
  \langle\Lambda^{0}|O_{-}^{s}| \Xi^{'0}_{c}\rangle=-\frac{2\sqrt{3}}{3}(X_{1}^{s}+3X_{2}^{s})(4\pi),\nonumber\\
   & \langle\Sigma^{0}|O_{-}^{d}|\Xi^{0}_{c}\rangle=-\frac{4\sqrt{3}}{3}X_{1}^{d}(4\pi),\hspace{1.0cm}
     \langle\Sigma^{0}|O_{-}^{s}|\Xi^{0}_{c}\rangle=-\frac{2\sqrt{3}}{3}(X_{1}^{s}+3X_{2}^{s})(4\pi),\nonumber\\
   &  \langle\Lambda^{0}|O_{-}^{d}|\Xi^{0}_{c}\rangle=-4X_{2}^{d}(4\pi),\hspace{1.7cm}
        \langle\Lambda^{0}|O_{-}^{s}|\Xi^{0}_{c}\rangle=-2(X_{1}^{s}-X_{2}^{s})(4\pi),
 \end{align}
and for DCS processes are\\
\begin{align}
 & \langle\Sigma^{0}|O_{-}^{D}|\Omega_{c}^{0}\rangle=\frac{4}{3}\sqrt{2}X_{1}^D(4\pi),\hspace{3cm}
  \langle\Lambda|O_{-}^{D}|\Omega_{c}^{0}\rangle=-4\sqrt{6}X_{2}^D(4\pi).
\end{align}
where we have introduced the bag integrals $X_1$ and $X_2$
\begin{align} \label{eq:X}
&X_1=\int^R_0 r^2 dr (u_s v_u-v_s u_u)(u_c v_d -v_c u_d), \quad
X_2=\int^R_0 r^2 dr (u_s u_u+v_s v_u)(u_c u_d +v_c v_d), \nonumber\\
&X_1^d= \int^R_0r^2 dr (u_uv_u-v_u u_u)(u_c v_u -v_c u_u),\quad
X_2^d= \int^R_0r^2 dr (u_u u_c+v_u v_c)(u_u u_u +v_u v_u),\nonumber\\
&X_1^s= \int^R_0r^2 dr (u_sv_u-v_s u_u)(u_c v_s -v_c u_s),\quad
X_2^s= \int^R_0r^2 dr (u_s u_u+v_s v_u)(u_c u_s +v_c v_s),\nonumber\\
&X_1^D= \int^R_0r^2 dr (u_uv_u-v_u u_u)(u_c v_s -v_c u_s),\quad
X_2^D= \int^R_0r^2 dr (u_u u_u+v_u v_u)(u_c u_s +v_c v_s),
\end{align}
with the numbers $X_1=3.56\times 10^{-6}, X_2=1.74\times 10^{-4},
X_1^d=0, X_2^d= 1.60\times 10^{-4}$ , $X_1^s= 2.60\times 10^{-6},
X_2^s=1.96\times 10^{-4}$, $X_1^D=0, X_2^D= 1.78\times 10^{-4}$.
To obtain numerical results, we have employed the following bag parameters
\be
m_u=m_d=0, \quad m_s=0.279~{\rm GeV}, \quad m_c=1.551~{\rm GeV}, \quad R=5~{\rm GeV}^{-1},
\en
where $R$ is the radius of the bag.

\section{Axial-vector form factors}
\label{app:c}
The axial-vector form factor in the static limit can be expressed in the bag model as
\begin{equation}
g^{A(P)}_{\mathcal{B}'\mathcal{B}}=\la\mathcal{B}'\uparrow | b_{q_1}^\dagger b_{q_2}\sigma_z|
\mathcal{B}\uparrow\ra \int d^3\bm{r}\left(u_{q_1}u_{q_2}-\frac13 v_{q_1}v_{q_2}\right).
\end{equation}
The relevant results are
\begin{align}
&g^{A(\overline{K}^{0})}_{\Xi_{c}^{0}\Omega_{c}^{0}}=-\frac{\sqrt{6}}{3}(4\pi Z_{2}),\qquad
g^{A(\overline{K}^{0})}_{\Xi_{c}^{'0}\Omega_{c}^{0}}=\frac{2\sqrt{2}}{3}(4\pi Z_{2}),\qquad
g^{A(\pi^{+})}_{\Xi^{-}\Xi^{0}}=-\frac{1}{3}(4\pi Z_{1}),\nonumber\\
&g^{A(K^{-})}_{\Sigma^{+}\Xi^{0}}=\frac{5}{3}(4\pi Z_{2}),\qquad\qquad
g^{A(\overline{K}^{0})}_{\Omega_{c}^{0}\Xi^{0}}=0,\qquad\qquad\qquad
g^{A(\pi^{0})}_{\Xi^{0}\Xi^{0}}=-\frac{1}{6}(4\pi Z_{1}),\nonumber\\
&g^{A(K^{0})}_{\Xi^{0}\Lambda^{0}}=\frac{\sqrt{6}}{6}(4\pi Z_{2}),\qquad\quad
g^{A(K^{0})}_{\Xi^{0}\Sigma^{0}}=-\frac{5\sqrt{2}}{6}(4\pi Z_{2}),\quad
g_{\Omega_{c}^{0}\Omega_{c}^{0}}^{A(\eta_{8})}=-\frac{4\sqrt{6}}{9}(4\pi Z_{1}),\nonumber\\
&g^{A(\eta_{8})}_{\Sigma^{0}\Sigma^{0}}=\frac{\sqrt{6}}{3}(4\pi Z_{1}),\qquad\qquad
g^{A(\eta_{8})}_{\Sigma^{0}\Lambda^{0}}=0,\qquad\qquad\qquad
g^{A(\eta_{8})}_{\Lambda^{0}\Sigma^{0}}=0,\nonumber\\
&g^{A(\eta_{8})}_{\Lambda^{0}\Lambda^{0}}=-\frac{\sqrt{6}}{3}(4\pi Z_{1}),\qquad\quad
g^{A(\pi^{+})}_{\Sigma^{-}\Sigma^{0}}=\frac{2\sqrt{2}}{3}(4\pi Z_{1}),\qquad
g^{A(\pi^{+})}_{\Sigma^{-}\Lambda^{0}}=\frac{\sqrt{6}}{3}(4\pi Z_{1}),\nonumber\\
&g^{A(K^{+})}_{\Xi^{-}\Sigma^{0}}=-\frac{5\sqrt{2}}{6}(4\pi Z_{2}),\qquad
g^{A(K^{+})}_{\Xi^{-}\Lambda^{0}}=-\frac{\sqrt{6}}{6}(4\pi Z_{2}),\qquad
g^{A(\pi^{-})}_{\Sigma^{+}\Sigma^{0}}=-\frac{2\sqrt{2}}{3}(4\pi Z_{1}),\nonumber\\
&g^{A(\pi^{-})}_{\Sigma^{+}\Lambda^{0}}=\frac{\sqrt{6}}{3}(4\pi Z_{1}),\qquad\quad
g^{A(\pi^{0})}_{\Sigma^{0}\Lambda^{0}}=\frac{\sqrt{3}}{3}(4\pi Z_{1}),\qquad\qquad
g^{A(\pi^{0})}_{\Sigma^{0}\Sigma^{0}}=0,\nonumber\\
&g^{A(\pi^{0})}_{\Lambda^{0}\Lambda^{0}}=0,\qquad\qquad\qquad\quad
g^{A(\pi^{0})}_{\Lambda^{0}\Sigma^{0}}=\frac{\sqrt{3}}{3}(4\pi Z_{1}),\qquad\qquad
g^{A(\overline{K}^{0})}_{\Sigma^{0}\Xi^{0}}=-\frac{5\sqrt{2}}{6}(4\pi Z_{2}),\nonumber\\
&g^{A(\overline{K}^{0})}_{\Lambda^{0}\Xi^{0}}=\frac{\sqrt{6}}{6}(4\pi Z_{2}),\qquad\quad
g^{A(K^+)}_{\Omega^- \Xi^0}=-\frac{2\sqrt{6}}{3} (4\pi Z_2),
\end{align}
where the auxiliary bag integrals are given by
\begin{equation}
Z_1=\int r^2 dr\left(u_u^2 -\frac13 v_u^2\right),\qquad
Z_2=\int r^2 dr \left(u_u u_s -\frac13 v_u v_s\right).
\end{equation}
Numerically,
$(4\pi) Z_1= 0.65$ and  $(4\pi)Z_2=0.71$.

\vskip0.2cm
\noindent \textbf{Note Added}.
SH and GM contribute equally and are co-first authors, while FX is corresponding author.



\begin{thebibliography}{999}


\bibitem{Tanabashi:2018oca}
  M.~Tanabashi {\it et al.} [Particle Data Group],
  Phys.\ Rev.\ D {\bf 98},  030001 (2018).

\bibitem{Solovieva:2008fw} 
  E.~Solovieva {\it et al.},
  Phys.\ Lett.\ B {\bf 672}, 1 (2009)
  doi:10.1016/j.physletb.2008.12.062
  [arXiv:0808.3677 [hep-ex]].


\bibitem{CroninHennessy:2000bz} 
  D.~Cronin-Hennessy {\it et al.} [CLEO Collaboration],
  Phys.\ Rev.\ Lett.\  {\bf 86}, 3730 (2001)
  doi:10.1103/PhysRevLett.86.3730
  [hep-ex/0010035].
  
\bibitem{Frabetti:1994dp} 
  P.~L.~Frabetti {\it et al.} [E687 Collaboration],
  Phys.\ Lett.\ B {\bf 338}, 106 (1994).
  doi:10.1016/0370-2693(94)91351-X
  
  
\bibitem{Aaij:2017nav} 
  R.~Aaij {\it et al.} [LHCb Collaboration],
  Phys.\ Rev.\ Lett.\  {\bf 118}, no. 18, 182001 (2017)
  doi:10.1103/PhysRevLett.118.182001
  [arXiv:1703.04639 [hep-ex]].
  
   
\bibitem{Yelton:2017qxg} 
  J.~Yelton {\it et al.} [Belle Collaboration],
  Phys.\ Rev.\ D {\bf 97}, no. 5, 051102 (2018)
  doi:10.1103/PhysRevD.97.051102
  [arXiv:1711.07927 [hep-ex]].

   
\bibitem{Cheng:2017ove} 
  H.~Y.~Cheng and C.~W.~Chiang,
  Phys.\ Rev.\ D {\bf 95}, no. 9, 094018 (2017)
  doi:10.1103/PhysRevD.95.094018
  [arXiv:1704.00396 [hep-ph]].
  


\bibitem{Agaev:2017jyt} 
  S.~S.~Agaev, K.~Azizi and H.~Sundu,
  EPL {\bf 118}, no. 6, 61001 (2017)
  doi:10.1209/0295-5075/118/61001
  [arXiv:1703.07091 [hep-ph]].



\bibitem{Huang:2017dwn} 
  H.~Huang, J.~Ping and F.~Wang,
  Phys.\ Rev.\ D {\bf 97}, no. 3, 034027 (2018)
  doi:10.1103/PhysRevD.97.034027
  [arXiv:1704.01421 [hep-ph]].

\bibitem{An:2017lwg} 
  C.~S.~An and H.~Chen,
  Phys.\ Rev.\ D {\bf 96}, no. 3, 034012 (2017)
  doi:10.1103/PhysRevD.96.034012
  [arXiv:1705.08571 [hep-ph]].

  
  
\bibitem{Wang:2017hej} 
  K.~L.~Wang, L.~Y.~Xiao, X.~H.~Zhong and Q.~Zhao,
  Phys.\ Rev.\ D {\bf 95}, no. 11, 116010 (2017)
  doi:10.1103/PhysRevD.95.116010
  [arXiv:1703.09130 [hep-ph]].


  
\bibitem{Aaij:2018dso} 
  R.~Aaij {\it et al.} [LHCb Collaboration],
  Phys.\ Rev.\ Lett.\  {\bf 121}, no. 9, 092003 (2018)
  doi:10.1103/PhysRevLett.121.092003
  [arXiv:1807.02024 [hep-ex]].
  
\bibitem{Link:2003nq} 
  J.~M.~Link {\it et al.} [FOCUS Collaboration],
  Phys.\ Lett.\ B {\bf 561}, 41 (2003)
  doi:10.1016/S0370-2693(03)00388-5
  [hep-ex/0302033].

\bibitem{Adamovich:1995pf} 
  M.~I.~Adamivich {\it et al.} [WA89 Collaboration],
  Phys.\ Lett.\ B {\bf 358}, 151 (1995)
  doi:10.1016/0370-2693(95)00979-U
  [hep-ex/9507004].

\bibitem{Frabetti:1995bi} 
  P.~L.~Frabetti {\it et al.} [E687 Collaboration],
  Phys.\ Lett.\ B {\bf 357}, 678 (1995).
  doi:10.1016/0370-2693(95)00941-D
  
  
\bibitem{Cheng:2018rkz} 
  H.~Y.~Cheng,
  JHEP {\bf 1811}, 014 (2018)
  doi:10.1007/JHEP11(2018)014
  [arXiv:1807.00916 [hep-ph]].

\bibitem{Yelton:2017uzv} 
  J.~Yelton {\it et al.} [Belle Collaboration],
  Phys.\ Rev.\ D {\bf 97}, no. 3, 032001 (2018)
  doi:10.1103/PhysRevD.97.032001
  [arXiv:1712.01333 [hep-ex]].


\bibitem{Ammar:2002pf} 
  R.~Ammar {\it et al.} [CLEO Collaboration],
  Phys.\ Rev.\ Lett.\  {\bf 89}, 171803 (2002)
  doi:10.1103/PhysRevLett.89.171803
  [hep-ex/0207078].

  

  
  
  
\bibitem{Korner:1992wi} 
  J.~G.~Korner and M.~Kramer,
  Z.\ Phys.\ C {\bf 55}, 659 (1992).
  doi:10.1007/BF01561305
  
  
  
\bibitem{Xu:1992sw} 
  Q.~P.~Xu and A.~N.~Kamal,
  Phys.\ Rev.\ D {\bf 46}, 3836 (1992).
  doi:10.1103/PhysRevD.46.3836
  
  
\bibitem{Cheng:1991sn}
  H.~Y.~Cheng and B.~Tseng,
  Phys.\ Rev.\ D {\bf 46}, 1042 (1992)
  Erratum: [Phys.\ Rev.\ D {\bf 55}, 1697 (1997)].
  doi:10.1103/PhysRevD.55.1697, 10.1103/PhysRevD.46.1042
  
  
  
\bibitem{Cheng:1993gf}
  H.~Y.~Cheng and B.~Tseng,
  Phys.\ Rev.\ D {\bf 48}, 4188 (1993)
  doi:10.1103/PhysRevD.48.4188
  [hep-ph/9304286].



\bibitem{Ivanov:1997ra} 
  M.~A.~Ivanov, J.~G.~Korner, V.~E.~Lyubovitskij and A.~G.~Rusetsky,
  Phys.\ Rev.\ D {\bf 57}, 5632 (1998)
  doi:10.1103/PhysRevD.57.5632
  [hep-ph/9709372].

  
\bibitem{Dhir:2015tja} 
  R.~Dhir and C.~S.~Kim,
  Phys.\ Rev.\ D {\bf 91}, no. 11, 114008 (2015)
  doi:10.1103/PhysRevD.91.114008
  [arXiv:1501.04259 [hep-ph]].
  
  
\bibitem{Zhao:2018zcb}
  Z.~X.~Zhao,
  Chin.\ Phys.\ C {\bf 42} (2018) no.9,  093101
  doi:10.1088/1674-1137/42/9/093101
  [arXiv:1803.02292 [hep-ph]].
  
\bibitem{Brown:1966zz} 
  L.~S.~Brown and C.~M.~Sommerfield,
  Phys.\ Rev.\ Lett.\  {\bf 16}, 751 (1966).
  doi:10.1103/PhysRevLett.16.751


\bibitem{Gronau:1972pj} 
  M.~Gronau,
  Phys.\ Rev.\ D {\bf 5}, 118 (1972)
  Erratum: [Phys.\ Rev.\ D {\bf 5}, 1877 (1972)].
  doi:10.1103/PhysRevD.5.118, 10.1103/PhysRevD.5.1877


  
\bibitem{Cheng:2018hwl}
  H.~Y.~Cheng, X.~W.~Kang and F.~Xu,
  Phys.\ Rev.\ D {\bf 97}, no. 7, 074028 (2018)
  doi:10.1103/PhysRevD.97.074028
  [arXiv:1801.08625 [hep-ph]].

\bibitem{Zou:2019kzq} 
  J.~Zou, F.~Xu, G.~Meng and H.~Y.~Cheng,
  Phys.\ Rev.\ D {\bf 101}, no. 1, 014011 (2020)
  doi:10.1103/PhysRevD.101.014011
  [arXiv:1910.13626 [hep-ph]].
  
  
\bibitem{Cheng:2020wmk} 
  H.~Y.~Cheng, G.~Meng, F.~Xu and J.~Zou,
  Phys.\ Rev.\ D {\bf 101}, no. 3, 034034 (2020)
  doi:10.1103/PhysRevD.101.034034
  [arXiv:2001.04553 [hep-ph]].


\bibitem{Ablikim:2019zwe}
  M.~Ablikim {\it et al.} [BESIII Collaboration],
  ``Measurements of Weak Decay Asymmetries of $\Lambda_c^+\to pK_S^0$, $\Lambda\pi^+$, $\Sigma^+\pi^0$, and $\Sigma^0\pi^+$,''
  Phys.\ Rev.\ D {\bf 100}, 072004 (2019)
  [arXiv:1905.04707 [hep-ex]].



\bibitem{Chau:1995gk}
  L.~L.~Chau, H.~Y.~Cheng and B.~Tseng,
  ``Analysis of two-body decays of charmed baryons using the quark diagram scheme,''
  Phys.\ Rev.\ D {\bf 54}, 2132 (1996)
  [hep-ph/9508382].


\bibitem{Korner:1978tc} 
  J.~G.~Korner, G.~Kramer and J.~Willrodt,
  Z.\ Phys.\ C {\bf 2}, 117 (1979).
  doi:10.1007/BF01474126








\bibitem{Zenczykowski:1993jm}
  P.~\.{Z}enczykowski,
  ``Nonleptonic charmed baryon decays: Symmetry properties of parity violating amplitudes,''
  Phys.\ Rev.\ D {\bf 50}, 5787 (1994).






  
  






\bibitem{Cheng:1995fe}
H.~Cheng and B.~Tseng,
``1/M corrections to baryonic form-factors in the quark model,''
Phys.\ Rev.\ D \textbf{53}, 1457 (1996)
[arXiv:hep-ph/9502391 [hep-ph]].


\bibitem{Buchalla:1995vs}
  G.~Buchalla, A.~J.~Buras and M.~E.~Lautenbacher,
  Rev.\ Mod.\ Phys.\  {\bf 68}, 1125 (1996)
  [hep-ph/9512380].


\bibitem{MIT}
  A.~Chodos, R.~L.~Jaffe, K.~Johnson and C.~B.~Thorn,
  ``Baryon Structure in the Bag Theory'',
  Phys.\ Rev.\ D {\bf 10}, 2599 (1974);
  T.~A.~DeGrand, R.~L.~Jaffe, K.~Johnson and J.~E.~Kiskis,
  ``Masses and Other Parameters of the Light Hadrons'',
  Phys.\ Rev.\ D {\bf 12}, 2060 (1975).




\bibitem{Kroll}
   T.~Feldmann, P.~Kroll and B.~Stech,
  Phys.\ Lett.\ B {\bf 449}, 339 (1999)
  [hep-ph/9812269];
  Phys.\ Rev.\ D {\bf 58}, 114006 (1998)
  [hep-ph/9802409].




\bibitem{PerezMarcial:1989yh}
  R.~Perez-Marcial, R.~Huerta, A.~Garcia and M.~Avila-Aoki,
  Phys.\ Rev.\ D {\bf 40}, 2955 (1989)
  Erratum: [Phys.\ Rev.\ D {\bf 44}, 2203 (1991)].
  doi:10.1103/PhysRevD.44.2203, 10.1103/PhysRevD.40.2955




\bibitem{Pervin:2006ie} 
  M.~Pervin, W.~Roberts and S.~Capstick,
  Phys.\ Rev.\ C {\bf 74}, 025205 (2006)
  doi:10.1103/PhysRevC.74.025205
  [nucl-th/0603061].
  
\bibitem{Gutsche:2018utw} 
  T.~Gutsche, M.~A.~Ivanov, J.~G.~Körner and V.~E.~Lyubovitskij,
  Phys.\ Rev.\ D {\bf 98}, no. 7, 074011 (2018)
  doi:10.1103/PhysRevD.98.074011
  [arXiv:1806.11549 [hep-ph]].






\end{thebibliography}
\end{document}